\documentclass[twocolumn,notitlepage,aps,prl,superscriptaddress,nofootinbib,10pt,showpacs]{revtex4-2}

\newif\ifarxi   \arxifalse

\def\minisection#1{\ifarxi\section*{#1}\else\par\vskip20pt\noindent{\it #1.}\ \fi\noindent}
\def\minisubsection#1{\ifarxi\subsection*{#1}\else\par\vskip20pt\noindent{\it #1.}\ \fi\noindent}

\usepackage{amsmath}
\usepackage{amssymb}
\usepackage{graphicx}
\usepackage{hyperref}
\usepackage{color}
\usepackage{xcolor}
\hypersetup{
	colorlinks,
	linkcolor={black!30!blue},
	citecolor={black!30!blue},
	urlcolor={black!30!blue}
}
\usepackage[T1]{fontenc}
\usepackage[utf8]{inputenc}
\usepackage{amsthm}
\usepackage{bbm}
\usepackage{mathtools}
	\mathtoolsset{centercolon}
\usepackage{xspace}
\usepackage{array}
\usepackage{ifthen}
\usepackage{tikz}
	\usetikzlibrary{arrows}
	\usetikzlibrary{arrows.meta}
	\usetikzlibrary{scopes}
	\usetikzlibrary{decorations.pathreplacing}
	\usetikzlibrary{decorations.markings}
	\usetikzlibrary{decorations.shapes}
	\usetikzlibrary{shapes,arrows,positioning,patterns}
	\usetikzlibrary{matrix}
	\usetikzlibrary{fit}
	\usetikzlibrary{fadings}
	\usetikzlibrary{calc}
	\usetikzlibrary{math}
	\tikzset{>=latex}

\newcommand{\idty}{{\leavevmode\rm 1\mkern -5.4mu I}}

\def\Cx{{\mathbb C}}
\def\Ir{{\mathbb Z}}

\def\ind{{\mathop{\rm ind}\nolimits}\,}
\def\HH{\mathcal H}

\def\brAAket#1#2#3{\langle#1\vert#2\vert#3\rangle}

\def\ket #1{\vert#1\rangle}

\def\abs#1{\vert#1\vert}
\def\basis#1#2{\ket{#1,#2}}

\tikzset{decorate sep/.style 2 args=
	{decorate,decoration={shape backgrounds,shape=circle,shape size=#1,shape 
			sep=#2}}}
\def\punktabstand{0.1}
\def\punktlaenge{0.75}
\def\boxsize{1.3}
\def\boundingboxsize{2.6}
\def\opacityvalue{0.4}

\newcommand{\shift}{\begin{tikzpicture}[baseline={($ (current bounding box.center) + (0,.05) $)},scale=.7]
		\def\xdist{1.3}
		\def\ydist{.4}
		\tikzstyle{dot} =[circle,fill,inner sep = 1.8]
		\tikzstyle{rects} = [rectangle, rounded corners=6, very thick, draw=red, inner sep=1.6mm]
		\tikzstyle{arr} = [->,very thick,black]
		\foreach \y in {0,1,2}{		
			\node[dot] (0\y) at (\xdist*0,+\ydist-\ydist*\y) {};
		}
		\foreach \y in {0,1}{		
			\node[dot] (1\y) at (\xdist*1,+\ydist-\ydist*\y) {};
		}
		\foreach \y in {0,1,2,3}{		
			\node[dot] (2\y) at (\xdist*2,+\ydist-\ydist*\y) {};
		}
		\foreach \y in {0,1,2}{		
			\node[dot] (3\y) at (\xdist*3,+\ydist-\ydist*\y) {};
		}		
		\node[rects,fit=(00) (02)] {};
		\node[rects,fit=(10) (11)] {};
		\node[rects,fit=(20) (23)] {};
		\node[rects,fit=(30) (32)] {};
		\draw[dotted,thick] (-\xdist,0) -- (-.5*\xdist,0);
		\draw[dotted,thick] (3.5*\xdist,0) -- (4*\xdist,0);
		\node[inner sep=0,minimum width=.15cm] (leftup) at (-\xdist,\ydist) {};
		\node[inner sep=0,minimum width=.15cm] (rightup) at (4*\xdist,\ydist) {};
		\draw[arr] (leftup) to[out=30,in=150] (00);
		\draw[arr] (00) to[out=30,in=150] (10);
		\draw[arr] (10) to[out=30,in=150] (20);
		\draw[arr] (20) to[out=30,in=150] (30);
		\draw[arr] (30) to[out=30,in=150] (rightup);
	\end{tikzpicture}
}

\newcommand{\Umat}[1]{
	\left (
	\begin{tikzpicture}[baseline={($ (current bounding box.center) + (.5cm,-.1cm) $)},scale=#1]
		\draw[white] (-\boundingboxsize-.3,-\boundingboxsize-.3) rectangle (\boundingboxsize+.3,\boundingboxsize+.3);
		\coordinate (c) at (0,0);
		\draw ($ (-\boundingboxsize,\boundingboxsize) + (\boxsize,0) $) -- ($ (\boundingboxsize,-\boundingboxsize) + (0,\boxsize) $);
		\draw ($(-\boundingboxsize,\boundingboxsize) + (0,-\boxsize) $) -- ($ (\boundingboxsize,-\boundingboxsize) + (-\boxsize,0) $);
		\fill[lightgray] ($ (-\boundingboxsize,\boundingboxsize) + (\boxsize,0) $) -- ($ (\boundingboxsize,-\boundingboxsize) + (0,\boxsize) $) -- ($ (\boundingboxsize,-\boundingboxsize) + (-\boxsize,0) $) -- ($(-\boundingboxsize,\boundingboxsize) + (0,-\boxsize) $) -- cycle;			
		\draw[decorate sep={0.5pt}{\punktabstand cm},fill]  (-\boundingboxsize,\boundingboxsize) --  +(.3,-.3);
		\draw[decorate sep={0.5pt}{\punktabstand cm},fill] (\boundingboxsize,-\boundingboxsize) --  +(-.3,.3);
		\draw[gray,dashed,opacity=\opacityvalue] (0,\boundingboxsize) -- (0,-\boundingboxsize);
		\draw[gray,dashed,opacity=\opacityvalue] (\boundingboxsize,0) -- (-\boundingboxsize,0);
	\end{tikzpicture}
	\right )
}

\newcommand{\Udecmat}[1]{
	\left (
	\begin{tikzpicture}[baseline={($ (current bounding box.center) + (.5cm,-.1cm) $)},scale=#1]
		\draw[white] (-\boundingboxsize-.3,-\boundingboxsize-.3) rectangle (\boundingboxsize+.3,\boundingboxsize+.3);
		\coordinate (c) at (0,0);
		\draw ($ (-\boundingboxsize,\boundingboxsize) + (\boxsize,0) $) -- ($ (c) + (0,\boxsize)$) -- (c) -- ($ (c) + (-\boxsize,0) $) -- ($ (-\boundingboxsize,\boundingboxsize) + (0,-\boxsize) $);
		\fill[lightgray] ($ (-\boundingboxsize,\boundingboxsize) + (\boxsize,0) $) -- ($ (c) + (0,\boxsize)$) -- (c) -- ($ (c) + (-\boxsize,0) $) -- ($ (-\boundingboxsize,\boundingboxsize) + (0,-\boxsize) $) -- cycle;
		\draw ($ (\boundingboxsize,-\boundingboxsize) + (-\boxsize,0) $) -- ($ (c) + (0,-\boxsize)$) -- (c) -- ($ (c) + (\boxsize,0) $) -- ($ (\boundingboxsize,-\boundingboxsize) + (0,\boxsize) $);
		\fill[lightgray] ($ (\boundingboxsize,-\boundingboxsize) + (-\boxsize,0) $) -- ($ (c) + (0,-\boxsize)$) -- (c) -- ($ (c) + (\boxsize,0) $) -- ($ (\boundingboxsize,-\boundingboxsize) + (0,\boxsize) $) -- cycle;
		\draw[decorate sep={0.5pt}{\punktabstand cm},fill]  (-\boundingboxsize,\boundingboxsize) --  +(.3,-.3);
		\draw[decorate sep={0.5pt}{\punktabstand cm},fill] (\boundingboxsize,-\boundingboxsize) --  +(-.3,.3);
		\draw[gray,dashed,opacity=\opacityvalue] (0,\boundingboxsize) -- (0,-\boundingboxsize);
		\draw[gray,dashed,opacity=\opacityvalue] (\boundingboxsize,0) -- (-\boundingboxsize,0);
	\end{tikzpicture}
	\right )
}
\newcommand{\Vdecmat}[1]{
	\left (
	\begin{tikzpicture}[baseline={($ (current bounding box.center) + (.5cm,-.1cm) $)},scale=#1]
		\draw[white] (-\boundingboxsize-.3,-\boundingboxsize-.3) rectangle (\boundingboxsize+.3,\boundingboxsize+.3);
		\coordinate (c) at (0,0);
		\draw[fill=lightgray] ($ (c) +(-0.5*\boxsize,0.5*\boxsize) $) rectangle ++(\boxsize,-\boxsize);
		\node at ($ (c) +(-1.1*\boxsize,1.1*\boxsize) $) {\tiny$1$};
		\node at ($ (c) +(-0.75*\boxsize,0.75*\boxsize) $) {\tiny$1$};
		\node at ($ (c) +(1.1*\boxsize,-1.1*\boxsize) $) {\tiny$1$};
		\node at ($ (c) +(0.75*\boxsize,-0.75*\boxsize) $) {\tiny$1$};
		\draw[decorate sep={0.5pt}{\punktabstand cm},fill] (-\boundingboxsize,\boundingboxsize) -- +(\punktlaenge,-\punktlaenge);
		\draw[decorate sep={0.5pt}{\punktabstand cm},fill] (\boundingboxsize,-\boundingboxsize) -- +(-\punktlaenge,\punktlaenge);
		\draw[gray,dashed,opacity=\opacityvalue] (0,\boundingboxsize) -- (0,-\boundingboxsize);
		\draw[gray,dashed,opacity=\opacityvalue] (\boundingboxsize,0) -- (-\boundingboxsize,0);
	\end{tikzpicture}
	\right )	
}

\newcommand{\Vblocks}[1]{
	\left (
	\begin{tikzpicture}[baseline={($ (current bounding box.center) + (.5cm,-.1cm) $)},scale=#1]
		\draw[white] (-\boundingboxsize-.3,-\boundingboxsize-.3) rectangle (\boundingboxsize+.3,\boundingboxsize+.3);
		\coordinate (c) at (0,0);
		\draw[fill=lightgray] ($ (c) -(-0.5*\boxsize,0.5*\boxsize) $) rectangle ++(-\boxsize,\boxsize) rectangle ++(-\boxsize,\boxsize);
		\draw[fill=lightgray] ($ (c) +(0.5*\boxsize,-0.5*\boxsize) $) rectangle ++(\boxsize,-\boxsize);
		\draw[decorate sep={0.5pt}{\punktabstand cm},fill]  ($(c)+(-1.5*\boxsize,1.5*\boxsize)$)  -- (-\boundingboxsize,\boundingboxsize);
		\draw[decorate sep={0.5pt}{\punktabstand cm},fill] (\boundingboxsize,-\boundingboxsize) --  ($(c)+(1.5*\boxsize,-1.5*\boxsize)$);
		\draw[gray,dashed,opacity=\opacityvalue] (0,\boundingboxsize) -- (0,-\boundingboxsize);
		\draw[gray,dashed,opacity=\opacityvalue] (\boundingboxsize,0) -- (-\boundingboxsize,0);
	\end{tikzpicture}
	\right )	
}

\newcommand{\Ublocks}[1]{
	\left (
	\begin{tikzpicture}[baseline={($ (current bounding box.center) + (.5cm,-.1cm) $)},scale=#1]
		\draw[white] (-\boundingboxsize-.3,-\boundingboxsize-.3) rectangle (\boundingboxsize+.3,\boundingboxsize+.3);
		\coordinate (c) at (0,0);
		\draw[fill=lightgray] (c) rectangle ++(-\boxsize,\boxsize) rectangle ++(-\boxsize,\boxsize);
		\draw[fill=lightgray] (c) rectangle ++(\boxsize,-\boxsize) rectangle ++(\boxsize,-\boxsize);
		\draw[gray,dashed,opacity=\opacityvalue] (0,\boundingboxsize) -- (0,-\boundingboxsize);
		\draw[gray,dashed,opacity=\opacityvalue] (\boundingboxsize,0) -- (-\boundingboxsize,0);
	\end{tikzpicture}
	\right )	
}

\newcommand{\littlebox}[3]{\fill[black
	] ($ (tl) +(#1*#3-#1,-#1*#2+#1) $) rectangle ++(#1,-#1);}

\newcommand{\littleboxgray}[3]{\fill[lightgray
	] ($ (tl) +(#1*#3-#1,-#1*#2+#1) $) rectangle ++(#1,-#1);}

\newcommand{\Mpicgraydemo}[4]{
	\left (
	\begin{tikzpicture}[baseline={($ (current bounding box.center) + (.5cm,-.1cm) $)},scale=#1]
		\draw[white] (-\boundingboxsize-.3,-\boundingboxsize-.3) rectangle (\boundingboxsize+.3,\boundingboxsize+.3);
		\coordinate (c) at (0,0);
		\coordinate (tl) at (-\boundingboxsize,\boundingboxsize);
		\pgfmathsetmacro{\DD}{#2}
		\pgfmathsetmacro{\nn}{#3}
		\pgfmathsetmacro{\mm}{#4}
		\pgfmathsetmacro{\boxwidth}{2*\boundingboxsize/\DD}
		\littlebox{\boxwidth}{\nn}{\nn}
		\littlebox{\boxwidth}{\nn}{\mm}
		\littlebox{\boxwidth}{\mm}{\nn}
		\littlebox{\boxwidth}{\mm}{\mm}
		\fadenkreuzdemo
	\end{tikzpicture}
	\right )	
}

\newcommand{\fadenkreuznew}{
	\colorlet{fadenkreuzcolor}{red}
	\draw[fadenkreuzcolor,line width=0pt] (-.333333*\boundingboxsize,\boundingboxsize) -- (-0.333333*\boundingboxsize,-\boundingboxsize);
	\draw[fadenkreuzcolor,line width=0pt] (.333333*\boundingboxsize,\boundingboxsize) -- (0.333333*\boundingboxsize,-\boundingboxsize);
	\draw[fadenkreuzcolor,line width=0pt] (\boundingboxsize,-.333333*\boundingboxsize) -- (-\boundingboxsize,-.333333*\boundingboxsize);
	\draw[fadenkreuzcolor,line width=0pt] (\boundingboxsize,.333333*\boundingboxsize) -- (-\boundingboxsize,.333333*\boundingboxsize);
}

\newcommand{\fadenkreuzdemo}{
	\colorlet{fadenkreuzcolor}{blue}
	\draw[fadenkreuzcolor,line width=0pt] (-\boundingboxsize,0) -- (\boundingboxsize,0);
	\draw[fadenkreuzcolor,line width=0pt] (0,-\boundingboxsize) -- (0,\boundingboxsize);
}

\newcommand{\Mpicgraydemoarrows}[6]{
	\left (
	\begin{tikzpicture}[baseline={($ (current bounding box.center) + (.5cm,-.1cm) $)},scale=#1]
		\draw[white] (-\boundingboxsize-.3,-\boundingboxsize-.3) rectangle (\boundingboxsize+.3,\boundingboxsize+.3);
		\coordinate (c) at (0,0);
		\coordinate (tl) at (-\boundingboxsize,\boundingboxsize);
		\pgfmathsetmacro{\DD}{#2}
		\pgfmathsetmacro{\nn}{#3}
		\pgfmathsetmacro{\mm}{#4}
		\pgfmathsetmacro{\nnold}{#5}
		\pgfmathsetmacro{\mmold}{#6}
		\pgfmathsetmacro{\boxwidth}{2*\boundingboxsize/\DD}
		\littleboxgray{\boxwidth}{\nnold}{\nnold}
		\littleboxgray{\boxwidth}{\nnold}{\mmold}
		\littleboxgray{\boxwidth}{\mmold}{\nnold}
		\littleboxgray{\boxwidth}{\mmold}{\mmold}
		\littlebox{\boxwidth}{\nn}{\nn}
		\littlebox{\boxwidth}{\nn}{\mm}
		\littlebox{\boxwidth}{\mm}{\nn}
		\littlebox{\boxwidth}{\mm}{\mm}
		\ifthenelse{\nn=\nnold \AND \mm=\mmold}{}{
			\draw[red,thick,->] ($ (tl) + (\boxwidth*\nnold-0.5*\boxwidth,-\boxwidth*\mmold+0.5*\boxwidth) $) -- ($ (tl) + (\boxwidth*\nn-0.5*\boxwidth,-\boxwidth*\mm+0.5*\boxwidth) $);
			\draw[red,thick,->] ($ (tl) + (\boxwidth*\mmold-0.5*\boxwidth,-\boxwidth*\nnold+0.5*\boxwidth) $) -- ($ (tl) + (\boxwidth*\mm-0.5*\boxwidth,-\boxwidth*\nn+0.5*\boxwidth) $);
			\ifthenelse{\nn=\nnold}{
				\draw[red,thick,->] ($ (tl) + (\boxwidth*\mmold-0.5*\boxwidth,-\boxwidth*\mmold+0.5*\boxwidth) $) -- ($ (tl) + (\boxwidth*\mm-0.5*\boxwidth,-\boxwidth*\mm+0.5*\boxwidth) $);}{
				\draw[red,thick,->] ($ (tl) + (\boxwidth*\nnold-0.5*\boxwidth,-\boxwidth*\nnold+0.5*\boxwidth) $) -- ($ (tl) + (\boxwidth*\nn-0.5*\boxwidth,-\boxwidth*\nn+0.5*\boxwidth) $);
			}
		}
		\fadenkreuzdemo
	\end{tikzpicture}
	\right )	
}

\newcommand{\UblocksIdtyUtinyU}[1]{
	\left (
	\begin{tikzpicture}[baseline={($ (current bounding box.center) + (.5cm,-.1cm) $)},scale=#1]
		\draw[white] (-\boundingboxsize-.3,-\boundingboxsize-.3) rectangle (\boundingboxsize+.3,\boundingboxsize+.3);
		\coordinate (c) at (0,0);
		\draw[fill=lightgray] (c) rectangle ++(-\boxsize,\boxsize); 
		\draw[fill=white] ($(c) + (-\boxsize,\boxsize)$) rectangle ++(-\boxsize,\boxsize);
		\draw[fill=white] (c) rectangle ++(\boxsize,-\boxsize);
		\draw[fill=lightgray] ($(c) +(\boxsize,-\boxsize)$) rectangle ++(\boxsize,-\boxsize);
		\draw[gray,dashed,opacity=\opacityvalue] (0,\boundingboxsize) -- (0,-\boundingboxsize);
		\draw[gray,dashed,opacity=\opacityvalue] (\boundingboxsize,0) -- (-\boundingboxsize,0);
		\node at ($(c)+(-1.5*\boxsize,1.5*\boxsize)$) {\scalebox{.7}{$\idty$}};
		\node at ($(c)+(-.5*\boxsize,.5*\boxsize)$) {\scalebox{.70}{$W_{\text{-}1}$}};
		\node at ($(c)+(.5*\boxsize,-.5*\boxsize)$) {\scalebox{.70}{$\idty$}};
		\node at ($(c)+(1.5*\boxsize,-1.5*\boxsize)$) {\scalebox{.70}{$W_1$}};
	\end{tikzpicture}
	\right )	
}

\newcommand{\UblocksUIdtytinyU}[1]{
	\left (
	\begin{tikzpicture}[baseline={($ (current bounding box.center) + (.5cm,-.1cm) $)},scale=#1]
		\draw[white] (-\boundingboxsize-.3,-\boundingboxsize-.3) rectangle (\boundingboxsize+.3,\boundingboxsize+.3);
		\coordinate (c) at (0,0);
		\draw[fill=white] (c) rectangle ++(-\boxsize,\boxsize); 
		\draw[fill=lightgray] ($(c) + (-\boxsize,\boxsize)$) rectangle ++(-\boxsize,\boxsize);
		\draw[fill=lightgray] (c) rectangle ++(\boxsize,-\boxsize);
		\draw[fill=white] ($(c) +(\boxsize,-\boxsize)$) rectangle ++(\boxsize,-\boxsize);
		\draw[gray,dashed,opacity=\opacityvalue] (0,\boundingboxsize) -- (0,-\boundingboxsize);
		\draw[gray,dashed,opacity=\opacityvalue] (\boundingboxsize,0) -- (-\boundingboxsize,0);
		\node at ($(c)+(-1.5*\boxsize,1.5*\boxsize)$) {\scalebox{.70}{$W_{\text{-}2}$}};
		\node at ($(c)+(-.5*\boxsize,.5*\boxsize)$) {\scalebox{.70}{$\idty$}};
		\node at ($(c)+(.5*\boxsize,-.5*\boxsize)$) {\scalebox{.70}{$W_0$}};
		\node at ($(c)+(1.5*\boxsize,-1.5*\boxsize)$) {\scalebox{.70}{$\idty$}};
	\end{tikzpicture}
	\right )	
}

\newcommand{\Vthreeexamplenew}[1]{
	\left (
	\begin{tikzpicture}[baseline={($ (current bounding box.center) + (.5cm,-.1cm) $)},scale=#1]
		\draw[white] (-\boundingboxsize-.3,-\boundingboxsize-.3) rectangle (\boundingboxsize+.3,\boundingboxsize+.3);
		\coordinate (c) at (0,0);
		\coordinate (tl) at (-\boundingboxsize,\boundingboxsize);
		\pgfmathsetmacro{\boxwidth}{2*\boundingboxsize/6}
		\littleboxgray{\boxwidth}{1}{1};
		\littleboxgray{\boxwidth}{4}{4};
		\littlebox{\boxwidth}{2}{3};
		\littlebox{\boxwidth}{3}{2};
		\littlebox{\boxwidth}{5}{6};
		\littlebox{\boxwidth}{6}{5};
		\fadenkreuznew
	\end{tikzpicture}
	\right )	
}

\newcommand{\Uthreeexamplenewer}[1]{
	\left (
	\begin{tikzpicture}[baseline={($ (current bounding box.center) + (.5cm,-.1cm) $)},scale=#1]
		\draw[white] (-\boundingboxsize-.3,-\boundingboxsize-.3) rectangle (\boundingboxsize+.3,\boundingboxsize+.3);
		\coordinate (c) at (0,0);
		\coordinate (tl) at (-\boundingboxsize,\boundingboxsize);
		\pgfmathsetmacro{\boxwidth}{2*\boundingboxsize/6}
		\foreach \n in {1,2,3}{
			\foreach \m in {1,2,3}{
				\littlebox{\boxwidth}{\m}{\n};
				\littlebox{\boxwidth}{\m}{\n};
				\littlebox{\boxwidth}{\m}{\n};
			}
		}
		\foreach \n in {4,5,6}{
			\foreach \m in {4,5,6}{
				\littlebox{\boxwidth}{\m}{\n};
				\littlebox{\boxwidth}{\m}{\n};
				\littlebox{\boxwidth}{\m}{\n};
			}
		}		
		\fadenkreuznew
	\end{tikzpicture}
	\right )	
}

\newcommand{\Mpicgraynew}[4]{
	\left (
	\begin{tikzpicture}[baseline={($ (current bounding box.center) + (.5cm,-.1cm) $)},scale=#1]
		\draw[white] (-\boundingboxsize-.3,-\boundingboxsize-.3) rectangle (\boundingboxsize+.3,\boundingboxsize+.3);
		\coordinate (c) at (0,0);
		\coordinate (tl) at (-\boundingboxsize,\boundingboxsize);
		\pgfmathsetmacro{\DD}{#2}
		\pgfmathsetmacro{\nn}{#3}
		\pgfmathsetmacro{\mm}{#4}
		\pgfmathsetmacro{\boxwidth}{2*\boundingboxsize/\DD}
		\foreach \ii in {1,...,\DD}{
			\ifthenelse{\ii=\nn \OR \ii=\mm}{}{			\littleboxgray{\boxwidth}{\ii}{\ii}}
		}
		\littlebox{\boxwidth}{\nn}{\nn}
		\littlebox{\boxwidth}{\nn}{\mm}
		\littlebox{\boxwidth}{\mm}{\nn}
		\littlebox{\boxwidth}{\mm}{\mm}
		\fadenkreuznew
	\end{tikzpicture}
	\right )	
}

\begin{document}

\title{An algorithm to factorize quantum walks into shift and coin operations}
\author{C. Cedzich}
\affiliation{Quantum Technology Group, Heinrich Heine Universit\"at D\"usseldorf, Universit\"atsstr. 1, 40225 D\"usseldorf, Germany}
\author{T. Geib}
\affiliation{Institut f\"ur Theoretische Physik, Leibniz Universit\"at Hannover, Appelstr. 2, 30167 Hannover, Germany}
\author{R.~F. Werner}
\affiliation{Institut f\"ur Theoretische Physik, Leibniz Universit\"at Hannover, Appelstr. 2, 30167 Hannover, Germany}

\begin{abstract}
We provide an algorithm that factorizes one-dimensional quantum walks into a protocol of two basic operations: A fixed conditional shift that transports particles between cells and suitable coin operators that act locally in each cell. This allows to tailor quantum walk protocols to any experimental setup by rephrasing it on the cell structure determined by the experimental limitations. We give the example of a walk defined on a qutrit chain compiled to run an a qubit chain. 
\end{abstract}

\maketitle
	
\ifarxi\section*{Introduction}\else\fi

In this paper we demonstrate an algorithm that factorizes any one-dimensional quantum walk into a finite ``protocol'' of shift and coin operations within a given cell structure. 

Such a factorization is of fundamental importance for the understanding of single-particle dynamics in discrete time and space, and closes an important gap between two different perspectives. Only the two together give a complete understanding, and allow to decide whether a given task be achieved with available building blocks. The description of the tasks is by conditions ``from without''. The corresponding definition of a quantum walk in this perspective is that of a one-step unitary operator on a lattice system satisfying a locality condition. Many overarching results, like the topological classification of walks with symmetries \cite{TopClass,ShortVersion,UsOnTI,F2W} are based on just such an axiomatic characterization.
On the other hand, one may take a constructive approach to quantum walks, defining the class of quantum walks ``from within'' in terms of a few available operations. This is the natural approach for experimental implementation. Also the construction of explicit models as analogues of condensed matter systems \cite{Shikano:2010id,UsOnCantor,InhomogeneousWalkFillman,Sajid,KitaExploring,Asbo,cages} and the design of quantum walk based algorithms \cite{santha_QW_search,AmbainisKR05,Shenvi:2003be,PhysRevA.98.032115,ComprehensiveOverview} follows this approach. 

Only when the two approaches are demonstrably equivalent one is sure that no road blocks to implementation have been overlooked. The strongest way of showing this equivalence is a compilation algorithm, which produces from any abstractly given walk a factorization into a sequence of operations. In this paper we provide such a compilation method, breaking a general walk into two kinds of operations: Coins, rotating each cell separately and the conditional shift. The shift is specified by singling out one basis vector in each cell and then shifting only these components. The shift operation and the cell structure, with cells of dimension $2\leq d<\infty$, will be arbitrary but fixed throughout.

Clearly, every shift-coin protocol is a quantum walk according to the axiomatic definition. For the converse, only partial results exist. In the translation invariant setting, Fourier transformation maps any quantum walk to a finite dimensional unitary, which then can be factorized into shifts and coins via techniques that were originally developed for filter banks \cite{vogts2009discrete,gao2001factorization,meyer1996quantum}. Another known technique provides a factorization using a grouping of the system into sufficiently large cells \cite{OldIndex}. We actually use that technique below as one step of our algorithm, but additional work is required, when the shift capabilities and with it the cells are fixed. 

A shift-coin decomposition is required for a number of theoretical tasks. Firstly, to couple walks to gauge fields: Such fields are implemented as commutation phases of the shift operators \cite{ewalks,UsOnMag,locQuasiPer,ElektricExp}. 
Secondly, one would like to consider a walk as the one-particle sector of an interacting system, known abstractly as a quantum cellular automaton (QCA) \cite{Farrelly2020reviewofquantum,arrighi2019overview,schumacher2004reversible}. The best construction for this \cite{vogts2009discrete} uses exactly a decomposition as provided in our paper.  The analogy between walks and QCAs was a guiding idea in \cite{OldIndex}, so that one can hope for an extension of our results to the QCA setting.

Being able to factorize any local unitary is highly desirable also from an experimenter's perspective since shift and coin operations are implementable in various platforms such as neutral atoms in optical lattices \cite{OpticalLattice1}, trapped ions \cite{TrappedIons1,TrappedIons2}, light-pulses in optical fibres \cite{OpticalFibers1,OpticalFibers2} and photonic waveguide arrays \cite{WaveGuide1,WaveGuide2}.
On the other hand, our algorithm provides a concrete method to adapt a given quantum walk to the experimental set-up at hand. We demonstrate this by compiling a quantum walk with three-dimensional coins \cite{three-state-konno,three-state-temperature,Boettcher_2018} for systems where the experimental setup provides only qubit cells.

\minisection{The systems}

We consider the discrete time dynamics of single particles, so-called \emph{quantum walks}, on the 1D lattice $\Ir$ ``from without''. These systems are described by a unitary operator $W$ subject to a locality condition on an arbitrary but fixed cell structure
\begin{equation}\label{eq:hs}
	\HH=\bigoplus_{x\in\Ir}\HH_x,
\end{equation}
with uniformly bounded cell dimensions $2\leq d_x<N<\infty$. We denote the basis of such $\HH$ by $\basis xi$ with $x\in\Ir$ and $i=1,\dots,d_x$.
The locality condition is expressed as a finite upper bound $L<\infty$ on the interaction length, i.e.
\begin{equation}\label{eq:strict_locality}
	\abs{x-y}>L\quad\Rightarrow\quad\brAAket{x,i}{W}{y,j}=0.
\end{equation}
Clearly, this abstract point of view is independent of the given cell structure, and reorganizing the cells leads merely to a different but still finite interaction length.

Below we provide a factorization algorithm that allows us to compile any such \emph{banded unitary} $W$ as a quantum walk ``from within'', i.e. as a sequence
\begin{equation}\label{eq:USC}
	W=C_0S^{n_1}C_1\cdots S^{n_{i}}C_i
\end{equation} 
of two basic operations that are determined by the cell structure \eqref{eq:hs}: The conditional shift operator $S$ transports the first basis vector of each cell one cell to the right, i.e. 
\begin{equation}\label{eq:shift}
	S=\shift,
\end{equation}
and coin operators $C$ that act locally as a $d_x$-dimensional unitary $C(x)$.

\minisection{The Algorithm}

Given some banded unitary $W$, fix some cell structure according to \eqref{eq:hs}, thereby determining the interaction length $L$ in \eqref{eq:strict_locality}. Our factorization algorithm consists of five steps:

\minisubsection{\textnormal{\texttt{Step 0}} - deal with non-vanishing indices}

In this preparatory step we bring the walk into a standard form with zero net flow of information. This is measured by an integer-valued index $\ind(W)$  \cite{KITAEV20062,OldIndex}. Important for our purpose is that the index is additive, i.e.\ $\ind(W_1W_2)=\ind(W_1)+\ind(W_2)$. Moreover, $\ind(S)=1$ so that the walk $S^{-n}W$, with $n=\ind(W)$ has vanishing index. This allows us to henceforth assume that the walks we consider have vanishing net information flow, i.e. $\ind(W)=0$.

\minisubsection{\textnormal{\texttt{Step 1}} - decouple periodically}

Any walk $W$ on the integers with vanishing index can be decoupled into two half-line walks via a local decoupling \cite{OldIndex} , i.e. we can write $W$ as
\begin{equation}
	\Umat{.3}=\Vdecmat{.3}\Udecmat{.3},
\end{equation}
where the first operator on the right acts non-trivially only on a block of $2L$ cells and the second one is decoupled into a left and a right part.

Due to the locality of the decoupling we can repeat this periodically every $2L$ cells which gives
\begin{equation}\label{eq:decomp}
	\Umat{.3}=\Vblocks{.3}\Ublocks{.3},
\end{equation}
where the blocks on the right side are of different size since the local dimensions of the cells they entail varies. Moreover, it is important to note that these blocks act on overlapping subspaces.

If arbitrary shift operations were available, this would essentially solve our problem. The remaining steps are necessary to make do with just the given shift and cell structure.

\minisubsection{\textnormal{\texttt{Step 2}} - parametrize blocks by elementary unitaries}

Each block in \eqref{eq:decomp} is a $D\times D$-dimensional unitary with $D$ depending on the dimensions of the $2L$ cells in the block. Every such block can be parametrized as a product of  $D(D-1)/2$ ``elementary unitaries'' \cite{murnaghan1962unitary,ReckZeilinger,Huber} of the form 
\begin{equation}\label{eq:Mnm}
	M^{nm}=
	\begin{pmatrix}
		\idty & & & &  \\
		& a & & b&\\
		&  &  \idty  &  &\\
		& c & & d&\\
		&  &  &  &  \idty
	\end{pmatrix}
	\equiv\begin{bmatrix}\begin{pmatrix} a	&	b	\\	c	&	d	\end{pmatrix}\end{bmatrix}^{nm},
\end{equation}
which differ from the identity only in the four matrix elements at $(n,n),\allowbreak (n,m),\allowbreak (m,n)$ and $(m,m)$. These are replaced by the entries $a$, $b$, $c$ and $d$ of a unitary $2\times 2$ matrix. We indicate the positions of these elements by the superscript in the matrix $M$.
With this notation every block in \eqref{eq:decomp} can be written as
\begin{equation}\label{eq:decomposeunitary}
	\prod_{n=1}^{2D-1}\left(\prod_{m=n+1}^{2D}M^{nm}\right).
\end{equation}
Importantly, the algorithm of \cite{murnaghan1962unitary,ReckZeilinger,Huber} affects only the block itself, and therefore can be applied simultaneously in each block in \eqref{eq:decomp}.

\minisubsection{\textnormal{\texttt{Step 3}} - factorize elementary unitaries into shifts and coins}
	
The previous step reduces our task to writing each $M^{nm}$ as a shift-coin sequence on the given cell structure. To this end, we embed $M^{nm}$ into $\HH$ by padding it with identities on both sides, and take the indices $n,m$ as basis labels in the given cell structure, i.e.\ $n=\basis{x}{i}$ and $m=\basis{y}{j}$. 

We distinguish two cases: If $y=x$, $M^{nm}$ is already a coin. Otherwise, if $k=y-x\neq0$, the factorization of $M^{nm}$ requires the use of the conditional shift. First, we initialize the matrix elements of $M^{nm}$ in a coin 
\begin{equation}
	C_M=\begin{pmatrix}a	&	b	&	&	\\
		c	&	d	&	&	\\
		&	&	\idty
	\end{pmatrix}.
\end{equation}
at $y$. Then, we conjugate $C_M$ with $S^k$, which translates the basis element $\basis y1$ to the correct cell at $x$. Last, we conjugate with a coin $C$ that swaps basis elements in the cells at $x$ and $y$ according to
\begin{equation}
	C\basis xi=\basis x1 \quad\text{and}\quad C\basis yj=\basis y2.
\end{equation}
Combining these steps amounts to
\begin{equation}\label{eq:mnm_sc}
	M^{nm}=C^\dagger S^{-k}C_MS^{k}C,
\end{equation}
i.e.\ a sequence of shift and coin-operations specified by $k$, the swapping coin $C$ and the coin $C_M$ at $y$.

Let us illustrate the above by applying it to an elementary unitary $M^{13}$ on $\Cx^2\oplus\Cx^2$. Initializing $C_M$, shifting by $k=1$, and swapping with $C=\idty\oplus\sigma_1$ gives indeed
\begin{equation*}
	\Mpicgraydemo{0.2}434\stackrel{S}{\longrightarrow}\Mpicgraydemoarrows{0.2}41434\stackrel{C}{\longrightarrow}\Mpicgraydemoarrows{0.2}41314.
\end{equation*}

\minisubsection{\textnormal{\texttt{Step 4}} - assemble}

The last step is to apply \texttt{step 3} in parallel in each block in \eqref{eq:decomp}. In the special case of equal cells, each block is parametrized by elementary unitaries in the same way. Thus, we can use a common ``shift-skeleton'' for all blocks and we only have to adjust the coins $C_M$ and swaps $C$ in \eqref{eq:mnm_sc} in parallel in each block. 
In total, this gives a sequence of at most $5D(D-1)\sim\mathcal O(d^2L^2)$ coin and shift operations.

In the general case, the cell configuration varies from block to block, and with it the parametrization of each block into elementary unitaries. Thus, the above does not apply directly.
What helps us out in this case is the uniform bound $N$ on the local cell dimension which implies that there is only a finite number of different cell configurations of the blocks and thus only a finite number of different parametrizations \eqref{eq:decomposeunitary}. We further factorize the block diagonal matrices in \eqref{eq:decomp} such that each factor contains all blocks with the same block configuration.
For example, if there are two cell configurations that alternate we write the block diagonals in \eqref{eq:decomp} as
\begin{equation}\label{eq:ueven_uodd}
	\Ublocks{.3}=\UblocksIdtyUtinyU{.3}\UblocksUIdtytinyU{.3}.
\end{equation}
To each such sparse factor, we can apply \texttt{Step 3} with a common ``shift-skeleton'', each contributing $\mathcal O(D^2)$ shift and coin operations to the overall sequence.

This step is crucial: without it we could still find a shift-coin factorization for every block in \eqref{eq:decomp}, but without a common ``shift-skeleton'' we could not apply \texttt{Step 3} in parallel for all blocks which would lead to an infinite product.

This concludes our description of the algorithm.

\minisection{Example}

An important application of the above algorithm is to tailor a given quantum walk to another architecture that is determined by experimental constraints. As an example, we show how the so-called ``three-state'' quantum walk discussed in \cite{three-state-konno,Boettcher_2018,three-state-temperature} can be realized in a setup with qubit cells. As the name suggests, this walk is defined on a Hilbert space with three-dimensional cells as the shift-coin protocol
\begin{equation}
	W=S_1S_3^\dagger C
\end{equation}
where $S_1=S$ as in \eqref{eq:shift} and $S_3$ shifts the third basis vector in each cell. As coin we choose for all $x$ the three-dimensional Grover matrix $C(x)=\tfrac13\left(\begin{smallmatrix}	-1	&	2	&	2	\\	2 & -1 & 2	\\ 2 & 2 & -1 \end{smallmatrix}\right)$, but the following analysis applies with appropriate changes for arbitrary choices of the local coins. The index of $W$ vanishes by $\ind(W)=\ind(S_1)+\ind(S_3^\dagger)=\ind(S_1)-\ind(S_3)=0$, such that we can directly start with \texttt{Step 1}. 
To decouple $W$ we capitalize on the hemiolic relation between the old and the new cell structure, where two three-dimensional cells are interpreted as three two-dimensional cells, i.e.
\begin{equation*}
	\includegraphics[scale=0.13]{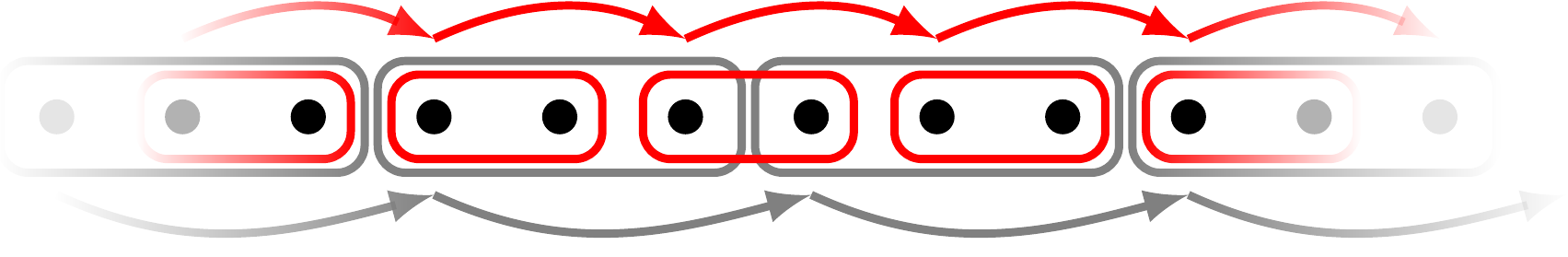}
\end{equation*}

On the new $\Cx^2$-cells $W$ has interaction length $L=2$ such that \texttt{Step 1} gives a decoupling every $2L=4$ \emph{new} cells. 
However, by the hemiolic relation between the old and the new cell structure we can also decouple every $2L$ cells in the \emph{old} cell structure where $L=1$. It thus suffices to consider blocks that contain only 3 instead of 4 new cells. 
One possible periodic decoupling according to \eqref{eq:decomp} results in the blocks
\begin{equation}\label{eq:Vblocknew}
	\scalebox{.85}{$\begin{pmatrix}
			1& & & \\
			&\sigma_1& & \\
			& &1& \\
			& & &\sigma_1\\
		\end{pmatrix}$}=\Vthreeexamplenew{0.2}=[\sigma_1]^{23}[\sigma_1]^{56}
\end{equation}
acting on $\HH_{x-1}\oplus\HH_{x}\oplus\HH_{x+1}$ ($x\in 3\Ir$) with respect to the $\Cx^2$-cells, and 

\begin{equation}\label{eq:Ublocknew}
	\frac13\scalebox{.85}{$\begin{pmatrix}
			2& 2& -1& & & \\
			2& -1& 2& & & \\
			-1&2& 2& & & \\
			& & & 2& 2& -1\\
			& & & 2& -1& 2\\
			& & & -1& 2& 2
		\end{pmatrix}$}
	=\Uthreeexamplenewer{0.2},
\end{equation}
acting  $\HH_x\oplus\HH_{x+1}\oplus\HH_{x+2}$.

The parametrization by elementary unitaries of the first block can be read off directly. 
Applying the algorithm of \cite{Huber,murnaghan1962unitary,ReckZeilinger} the second block is parametrized by the elementary unitaries $M^{12}=[H]^{12}$, $M^{14}=[A]^{14}$ and $M^{24}=[\sigma_1H]^{24}$ for the first $3\times3$ block, and analogously 
$M^{45}=[H]^{45}$, $M^{46}=[A]^{46}$ and  $M^{56}=[\sigma_1H]^{56}$ for the second $3\times3$ block, where $H=(\sigma_1+\sigma_3)/\sqrt{2}$ is the Hadamard coin and 
\begin{equation}
	A= \frac13\begin{pmatrix}
		2\sqrt2&1\\-1&2\sqrt2
	\end{pmatrix}.
\end{equation}
Hence, the blocks \eqref{eq:Ublocknew} are parametrized as 

\begin{widetext}
	\begin{equation}\label{eq:prodofelemunit}
		\Uthreeexamplenewer{0.2}=
		\stackrel{H}{\Mpicgraynew{0.2}612}
		\stackrel{A}{\Mpicgraynew{0.2}613}
		\stackrel{\sigma_1H}{\Mpicgraynew{0.2}623}
		\stackrel{H}{\Mpicgraynew{0.2}645}
		\stackrel{A}{\Mpicgraynew{0.2}646}
		\stackrel{\sigma_1H}{\Mpicgraynew{0.2}656}.
	\end{equation}
\end{widetext}
Thus, in \texttt{Step 3} we only need to consider $[\sigma_1]^{23}$ on $\HH_{x-1}\oplus\HH_{x}\oplus\HH_{x+1}$ and $[A]^{13}$, $[\sigma_1 H]^{23}$, $[H]^{45}$ and $[A]^{46}$ on $\HH_{x}\oplus\HH_{x+1}\oplus\HH_{x+2}$, since the remaining elementary unitaries are already coins. 
$[\sigma_1]^{23}$ is realized by $C_M=\sigma_1$ at $x$, $k=1$ and swapping coins at $x-1$ and $x$. 
The shift-coin factorizations of the  elementary unitaries in \eqref{eq:prodofelemunit}, with $k=y-x\neq0$ are parametrized by the following data:
\begin{center}
	\begin{tabular}{l|crc}
		&	$C_M(x)$ 		&   $k$	&	$C(x)$			\\\hline\\[-3mm]
		$[A]^{13}$	&	$A(x+1)$		&	$1$	&	$\sigma_1(x+1)$	\\
		$[A]^{46}$	&	$A(x+2)$		&	$1$	&	$\sigma_1(x+1)$	\\
		$[\sigma_1H]^{23}$	&	$H\sigma_1(x)$ & $-1$	&	$\idty$	\\
		$[H]^{45}$	&	$\sigma_1H\sigma_1(x+1)$ & $-1$	&	$\idty$
	\end{tabular}
\end{center}

Since the cells have constant dimension, in \texttt{Step 4} these block-factorisations can be performed in parallel with a common ``shift-skeleton''.

\minisection{Summary and Outlook}

We provided a concrete algorithm to factorize any one-dimensional quantum walk into a finite product of shift and coin operations on any given cell structure. This closes a long-standing gap in the understanding of such systems, but also has practical implications: on the one hand, it allows to adapt a given walk to any experimental set-up and, on the other, to either optimize with respect to the cell dimensions or the interaction length.

An interesting direction for future work is to optimize the length of the shift-coin protocols. One option that jumps to the eye is to homogenize a given cell structure by ``filling up'' each cell by locally adding innocent bystanders until all cells have the same dimension. This, however, would violate the assumption of a given fixed cell structure.

\minisection{Acknowledgements}
T. Geib and R. F. Werner acknowledge support from the DFG through SFB 1227 DQ-mat.

\bibliography{BIB}
\end{document}